\documentclass[apj,numberedappendix]{emulateapj}
\usepackage{natbib,apjfonts}

\newcommand{\be}{\begin{eqnarray}}
\newcommand{\ee}{\end{eqnarray}}
\newcommand{\beq}{\begin{equation}}
\newcommand{\eeq}{\end{equation}}
\def\lp{\left(}
\def\rp{\right)}

\slugcomment{Accepted for publication in The Astrophysical Journal}
\shorttitle{SURFACE MODES ON BURSTING NEUTRON STARS}
\shortauthors{PIRO \& BILDSTEN}

\begin{document}

\epsscale{1.2}


\title{Surface Modes on Bursting Neutron Stars and X-ray Burst Oscillations}

\author{Anthony L. Piro}
\affil{Department of Physics, Broida Hall, University of California,
	\\ Santa Barbara, CA 93106; piro@physics.ucsb.edu}

\and

\author{Lars Bildsten}
\affil{Kavli Institute for Theoretical Physics and Department of Physics,
Kohn Hall, University of California,
	\\ Santa Barbara, CA 93106; bildsten@kitp.ucsb.edu}


\begin{abstract}
  Accreting neutron stars (NSs) often show coherent modulations
during type I X-ray bursts, called burst oscillations. We consider
whether a nonradial mode can serve as an explanation for burst oscillations
from those NSs that are not magnetic. We find
that a surface wave in the shallow burning layer changes to a
crustal interface wave as the envelope cools, a new and previously
uninvestigated phenomenon. The surface modulations decrease
dramatically as the mode switches, explaining why burst oscillations
often disappear before burst cooling ceases. When we include rotational
modifications, we find mode frequencies and drifts consistent with
those observed. The large NS spin ($\approx270-620\ {\rm Hz}$)
needed to make this match implies that accreting NSs are spinning
at frequencies $\approx4\ {\rm Hz}$ above the burst oscillation. Since
the long-term stable asymptotic frequency is set by the crustal interface wave,
the observed late time frequency drifts are a probe of the composition and
temperature of NS crusts. We compare our model with the observed drifts
and persistent luminosities of X-ray burst sources, and find that NSs with
a higher average accretion rate show smaller drifts, as we predict. Furthermore,
the drift sizes are consistent with crusts composed of iron-like nuclei,
as expected for the ashes of the He-rich bursts that are exhibited
by these objects.
\end{abstract}




\section{Introduction}

  Type I X-ray bursts are the result of unstable nuclear burning on
the surface of accreting neutron stars (NSs)
\citep[see reviews by][]{bil98,sb03}, triggered by the extreme
temperature sensitivity of triple-$\alpha$ reactions
\citep{hv75,wt76,mc77,jos77,ll78}. They have rise times of seconds
with decay times ranging from tens to hundreds of seconds,
depending on the composition of the burning material. When the
NS is actively accreting the bursts repeat every few hours to days,
the timescale to accumulate an unstable column of fuel.

  Oscillations are often seen in the burst light curves both before and
after the burst peak \citep[][and references therein]{mun01}. They
have frequencies of $270-620\ {\rm Hz}$
\citep[with one case of $45\ {\rm Hz}$,][]{vs04} and
typically show positive drifts of $\approx1-5\textrm{ Hz}$.
During the burst rise, the frequency and amplitude evolution are consistent
with a hot spot from the burst ignition spreading over the NS surface
\citep{szs97}, but the positive drift in the decaying tail has not yet
been satisfactorily explained. \citet{cb00} explored
Strohmayer et al.'s (1997) hypothesis that this spin-up is simply
angular momentum conservation as the surface layers expand and
contract. \citet{hey00} and \citet{akl01} pointed out the importance of
general relativistic corrections, and \citet{cum02} eventually concluded
that this mechanism
underpredicts the observed drift sizes. These works did not resolve
the cause of the surface asymmetry at late times, long after any hot
spots should have spread over the surface \citep{bil95,slu02}. The
asymptotic frequency is
characteristic to a given object and is very stable over many
observations \citep[within 1 part in $10^3$,][]{mun02}. This fact, along with
burst oscillations seen from two accreting millisecond pulsars at their
non-bursting pulsar frequency (SAX J1808.4-3658, Chakrabarty et al. 2003;
XTE J1814-338, Strohmayer et al. 2003), have led many to conclude that
burst oscillations exactly indicate the NS spin frequency.

  However, it remains a mystery as to what creates the surface
asymmetry in those accreting NSs that do not show pulsations in their
persistent emission. In Table 1 we summarize the burst oscillations
seen from 12 non-pulsar NSs. Since these objects have weaker
magnetic fields than the accreting pulsars, their burst oscillations
may well be due to a different mechanism. This hypothesis is supported
by many differences between the burst oscillations from pulsars and
non-persistently pulsating NSs. The non-pulsars only show burst oscillations
in short ($2-10\textrm{ s}$) bursts (excluding superbursts, see Table 1),
while the pulsars have also shown burst oscillations in longer
bursts \citep[in XTE J1814-338,][]{str03}. The non-pulsars show
frequency drifts during
burst cooling, often late into the burst tail, while the pulsars only show
drifts during the burst rise with no frequency evolution after the burst
peak \citep{cha03,str03}. The non-pulsars have burst oscillations that are
highly sinusoidal while the pulsars show slight harmonic content
(compare the results of Muno, \"{O}zel, \& Chakrabarty 2002 with
Strohmayer et al. 2003). Finally, the pulsed amplitude as a function of
energy is different between the two categories of objects
\citep{moc03,ws04}.

\subsection{Nonradial Modes as Burst Oscillations}

  An attractive explanation for the burst oscillations in non-magnetic
NSs is that they originate in nonradial oscillations \citep{hey04}
since this is an obvious way to make large scale asymmetries in a liquid.
Nonradial oscillations on bursting NSs were previously studied by
\citet{mt87}, but this was before the discovery of burst oscillations
\citep{str96}. They did not incorporate many important physical details
such as a fast NS spin and the crustal interface mode which are crucial
to our arguments.
The angular and radial eigenfunctions that are allowed for such a
mode are severely restricted by the many properties of burst oscillations.
\citet{hey04} identified that the angular structure must be given by
an $m=1$ buoyant {\it r}-mode. His arguments for this are as follows. The
highly sinusoidal nature of the oscillations \citep{moc02},
implies an azimuthal quantum number of $m=1$ or $m=-1$ for the
surface asymmetry. A mode with frequency $\omega$ in a frame
co-rotating with the stellar surface is seen by an
inertial observer to have a frequency
\be
	\omega_{\rm obs} = |m\Omega-\omega|,
	\label{eq:obs}
\ee
where $\Omega$ is the NS spin. Since the frequency of a surface
wave, $\omega$, decreases as the star cools,
the mode must be traveling retrograde to the spin ($m>0$) for a
positive frequency drift in the observer's frame. The fast
NS spin (which we argue is close to the burst oscillation
frequencies of $\approx270-620\ {\rm Hz}$) alters the latitudinal
eigenfunctions, so that the angular eigenfunctions
are no longer given by spherical harmonics, and $\omega$ is similarly
modified (as we summarize in \S \ref{sec:rotation},  also see Longuet-Higgins 1968;
Bildsten, Ushomirsky, \& Cutler 1996, hereafter BUC96; Piro \& Bildsten 2004).
Due to these effects, \citet{hey04} concluded that {\it r}-modes are ideally
suited to be burst oscillations because they have $m>0$, their
latitudinal
eigenfunctions span a wide region around the equator so that
they may be easier to observe, and the rotational modifications
to $\omega$ result in small drifts.

\begin{figure}
\epsscale{1.15}
\plotone{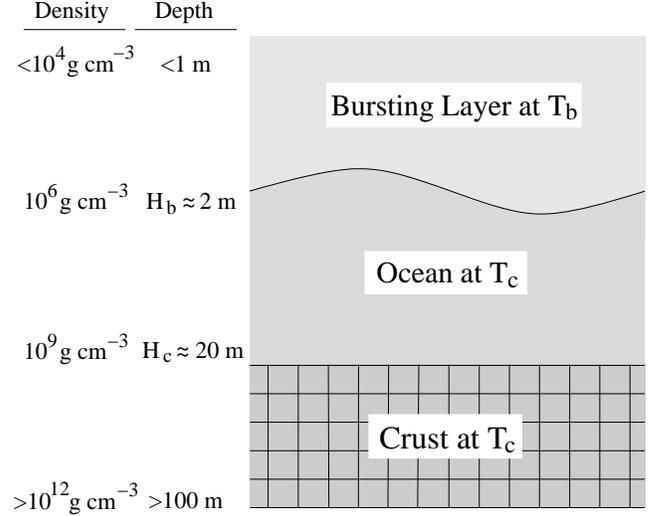}
\caption{Diagram of the NS surface layers during the
decaying tail of an X-ray burst. The surface is composed of three regions,
a hot layer of recently burned fuel with temperature $T_b$, a cooler ocean
with temperature $T_c$, and the outer crust also with $T_c$.
The bursting-layer/ocean interface is at a depth
determined by where unstable burning ignites during the X-ray burst, while
the ocean/crust interface occurs at the crystallization depth of the
ocean material. Each interface provides a buoyancy that is a natural place
for supporting modes. During the beginning of the burst the region
just above the bursting-layer/ocean interface is convective, but we ignore this
complication to simplify our analysis.}
\epsscale{1.2}
\label{fig:diagram}
\end{figure}

  The remaining outstanding problem with Heyl's explanation was in
identifying the correct radial structure of the mode. The outer NS
during burst cooling has three separate layers: a hot bursting layer
that was heated during the burning, a cooler ocean below, and finally
the shallow region of the crust (see Figure \ref{fig:diagram}). A natural
first guess for the cause of the burst oscillation is a shallow surface
wave excited in the hot bursting layer and riding the buoyancy at
the bursting-layer/ocean interface. One problem with this explanation
is that this mode overestimates the observed
frequency shifts (as the bursting layer cools from $10^9\textrm{ K}$ to
a few$\times10^8\textrm{ K}$, the mode frequency changes by too much,
even once rotational modifications are included). In addition, this mode
cannot reproduce the extreme stability of the measured asymptotic
frequencies (a surface wave's frequency will vary depending on the
temperature and resulting composition that is unique to each burst).

  We solve these difficulties by including both the shallow surface
wave {\it and} the crustal interface wave
\citep[][hereafter PB05]{pb05}. The latter is concentrated in the ocean
and rides the ocean/crust interface. These two modes have frequencies
close enough that they undergo an avoided crossing during the
burst cooling. The case we focus on here is one in which the energy in
the surface wave ``adiabatically'' changes into the crustal interface wave
as the burst cools. This switch solves the two problems listed
above because: (1) switching to a crustal interface wave ceases the frequency
evolution so that the predicted frequency drifts match the observed drifts,
and (2) the final frequency is stable and characteristic to each object because
it only depends on the crust's properties, which do not change from
burst to burst.

\subsection{Outline of Paper}

  In \S \ref{sec:sum} we use simple analytic arguments to describe our
idea for how the surface wave and crustal interface wave interact to reproduce
the main features of burst oscillations. In \S \ref{sec:env} we present time
dependent numerical models to simulate the NS surface during the cooling
following an X-ray burst. This model is used in \S \ref{sec:modes} to
calculate the eigenfrequency spectrum, focusing on the surface wave and
crustal interface wave, including rotational modifications
expected for a quickly rotating NS. We also estimate the
time dependent oscillation amplitude, showing that it decreases
as the surface wave changes into a crustal interface wave. In \S \ref{sec:drifts}
we compare the observed drifts and persistent luminosities with our analytic
mode estimates. We consider other modes in \S \ref{sec:othermodes}, and
examine the prospects for their detection. We then
conclude with a summary of our results in \S \ref{sec:concl}.


\section{Analytic Summary and Dependence on the Neutron Star Crust}
\label{sec:sum}

  We now outline why the shallow surface wave and crustal interface wave
are consistent with the properties of burst oscillations.
This demonstrates which properties of the NS
crust are constrained by burst oscillation observations.

  A shallow surface wave in the hot bursting layer (see Figure
\ref{fig:diagram}) has a frequency
\be
        \omega^2_{\rm s} = gH_bk^2\frac{\Delta\rho}{\rho},
	\label{eq:surface1}
\ee
where $g=GM/R^2$ is the surface gravitational acceleration,
$H_b=P/(\rho g)$ is the pressure scale height at the base of the
bursting layer, $k^2=\lambda/R^2$ is the transverse wavenumber,
$\Delta\rho/\rho=1-T_c\mu_b/(T_b\mu_c)$ is the fractional
density contrast at the burning depth, and $\mu_b$ ($\mu_c$) is
the mean molecular weight in the bursting layer (ocean). Throughout
we assume that the ocean and crust have the same temperature and
composition. In the non-rotating limit $\lambda=l(l+1)$,
where $l$ is the normal spherical harmonic quantum number. For a
quickly spinning NS, $\lambda$ is a function of $\Omega$ and
$\omega$, which can be accurately calculated using the ``traditional
approximation'' (as we summarize in \S \ref{sec:rotation}, or
see BUC96). In agreement with \citet{hey04} we show that the only low order,
rotationally modified angular mode that reproduces all of the properties of
the burst oscillations is a buoyant {\it r}-mode, which has
$\lambda\approx1/9\approx0.11$ in the quickly spinning limit
\citep[this mode is often identified as the $l=2$, $m=1$
{\it r}-mode from the slowly rotating limit,][]{pb04}.
Using the scalings from equation
(\ref{eq:surface1}),
\be
        \frac{\omega_{s}}{2\pi}
		=10.8\ {\rm Hz}\left(\frac{2Z_b}{A_b}\right)^{1/2}
			\left(\frac{T_b}{10^9\ {\rm K}}\right)^{1/2}
			\left(\frac{10\ {\rm km}}{R}\right)
			\nonumber
			\\
			\times \lp\frac{\lambda}{0.11}\rp^{1/2}
			\lp1-\frac{T_c}{T_b}
				\frac{\mu_b}{\mu_c}\rp^{1/2},
	\label{eq:surface2}
\ee
where $A_b$ and $Z_b$ are the mass number and charge of the ions in the
bursting layer, respectively, and we use an ideal gas equation of state.
The final term from the density discontinuity is $\approx0.8$ near the burst
peak. This frequency is consistent with the low-order modes found by
\citet{mt87}, but with $\lambda=2$, as appropriate for the $l=1$ modes on
a non-spinning NS that they studied.

  The crustal interface wave has a frequency set by the cool NS ocean (PB05),
\be
        \omega^2_c = \mu_0gH_ck^2,
	\label{eq:crust1}
\ee
where $\mu_0=\mu/P\sim10^{-2}$ (see \S \ref{sec:radial}) is the ratio
of the shear modulus to the pressure at the top of the crust, and the
scale height, $H_c$, is evaluated at this same depth. Equation
(\ref{eq:crust1}) is the dispersion relation for a shallow surface
wave, but with a factor of $\mu_0$ due to crust compression.
For the temperatures expected at the bottom
of the ocean, and using a pressure dominated by degenerate, relativistic
electrons (PB05)
\be
        \frac{\omega_c}{2\pi}
                =4.3\ {\rm Hz}\left(\frac{64}{A_c}\right)^{1/2}
		\left(\frac{T_c}{3\times10^8\ {\rm K}}\right)^{1/2}
	\nonumber
	\\
	\times
	\left(\frac{10\ {\rm km}}{R}\right)
	\lp\frac{\lambda}{0.11}\rp^{1/2},
	\label{eq:crust2}
\ee
where the prefactor is set to match our numerical results, and
$A_c$ is the mass number of nuclei in the crust.

  The shallow surface mode's frequency decreases as
the layer cools. Once $T_b$ cools to $T_c\approx3\times10^8\ {\rm K}$
(and using $\mu_b/\mu_c\approx1.0$), we find that the surface wave's
frequency drifts by $\approx9\ {\rm Hz}$, much larger than the
observed shifts. This is one of the key problems with explaining the burst
oscillations with only a shallow surface wave. When the crustal interface
wave is included, it is possible that during the cooling
$\omega_s=\omega_c$, at which point the surface wave evolves into a
crustal interface wave (a transition we discuss in
\S \ref{sec:timedependent}). The frequency then remains fixed
because $T_c$ does not change during the
X-ray burst. The drift then is only
$\Delta\omega=\omega_s-\omega_c$, where $\omega_s$ is its value at
the beginning of the burst when $T_b\approx10^9\ {\rm K}$,
resulting in $\Delta\omega/(2\pi)\lesssim5\ {\rm Hz}$, much closer
to observations. In this picture, the size of the drift is approximately set
by the composition and temperature of the crust. A crust that is
{\it hotter} or composed of {\it lighter} elements has a higher crustal
interface wave frequency, and therefore its drifts are {\it smaller}.
The observed drifts require a crust with $T_c=(2-6)\times10^8\ {\rm K}$
for $A_c=50-100$. We come back to these scalings when we consider
the observed drifts in \S \ref{sec:drifts}.


\section{Cooling X-ray Burst Envelope Models}
\label{sec:env}

  We numerically calculate the time evolution of the cooling surface layers and
ocean following \citet{cm04}. Since
the NS radius, $R\approx10\textrm{ km}$, is much greater than the pressure
scale height $H_c\approx(2-4)\times10^4\textrm{ cm}$, we approximate the surface
as having a constant gravitational acceleration,
$g=GM/R^2\approx2\times10^{14}\textrm{ cm s}^{-2}(M/1.4M_\odot)(10\ {\rm km}/R)^2$
(neglecting general relativity), and plane-parallel geometry. We use
$z$ and $x$ as our radial and transverse coordinates, respectively,
and in addition find it useful to use the column depth, denoted as
$y$ (defined by $dy=-\rho dz$), giving a pressure $P=gy$.

  The initial profile of the envelope following the burning is that of
a hot, flux dominated, bursting layer sitting above a cooler, lower flux ocean.
To mimic this situation, we assume a constant
flux of $F_b\approx10^{25}\textrm{ erg cm}^{-2}\textrm{ s}^{-1}$
(basically an Eddington flux as seen in radius expansion bursts that
often proceed burst oscillations) above a burning depth column of
$y_b=3\times10^8\textrm{ g cm}^{-2}$ \citep{bil98}. The choice of this depth
mainly affects the cooling timescale, with a deeper $y_b$ resulting in
an extended light curve. This has a small effect on the frequencies since
for an electron scattering dominated, constant flux envelope
$T_b\propto y_b^{1/4}$, so from equation (\ref{eq:surface2}),
$\omega_s\propto y_b^{1/8}$. Previous studies of constantly accreting NSs
have shown that the interior thermal balance is set by electron captures,
neutron emissions, and pycnonuclear reaction in the inner crust
\citep{mph90,zdn92,bb97,bb98,bro00,bro04} which release
$\approx1\textrm{ MeV}/m_p\approx10^{18}\textrm{ erg g}^{-1}$
\citep{hz90,hz03}. Depending on the accretion rate and thermal
structure of the crust, this energy will either be conducted into the
core or released into the ocean such that for an Eddington accretion rate
up to $\approx92\%$ of the energy is lost to the core
and exits as neutrinos \citep{bro00}. Since we wish to investigate how
the properties of the crust affect the characteristics of the burst oscillations
we treat this crustal flux, $F_c$ as a free variable, and consider its effects
around a value of $10^{21}\ {\rm erg\ cm^{-2}}\ {\rm s^{-1}}$
(as expected for $\dot{M}\approx10^{-9}M_\odot\textrm{ yr}^{-1}$,
about one-tenth the Eddington rate).

  We use the results of \citet{sch01} and \citet{woo04} to set the compositions.
Table 2 lists the key properties of our models, which sample a range of
conditions expected for the surfaces of bursting NSs. We consider only one nuclear
species per layer, where that species is chosen to represent the wide range
of elements that actually exist. In He-rich bursts, the burning
predominantly produces $\alpha$-elements and cooling lasts $\sim10\ {\rm s}$
as observed with burst oscillations. We expect more mixing during
these bursts due to convection at the ignition depth
\citep{woo04,wb05}, so that $\mu_b/\mu_c\sim1$. Models 1 and 2 are
meant to represent this type of burst, with Model 2 exploring the changes
that occur for a large $T_c$. On the other hand, bursts with a composition
closer to solar (mixed H/He bursts) produce a heavy ocean of ashes from
rp-process burning and also
less mixing so that $\mu_b/\mu_c\lesssim1$. These bursts occur on
objects such as GS $1826-24$ \citep{gal04} and have never shown
burst oscillations. Model 3 represents this case,
but without the extra heat source of the rp-process burning. This only
changes the timescale of the cooling and frequency evolution, and does
not affect the calculated mode frequencies nor drifts.

  The density of the liquid/solid transition, $\rho_c$,
when crystallization begins is set by the
dimensionless parameter
\begin{eqnarray}
        \Gamma &\equiv& \frac{(Z_ce)^2}{a_ck_{\rm B}T_c}
        \nonumber
        \\
        &=& 169\left(\frac{3}{T_{c,8}}\right)\left(\frac{Z_c}{30}\right)^2
        \left( \frac{64}{A_c}\right)^{1/3}
        \left(\frac{\rho}{10^9\textrm{ g cm}^{-3}} \right)^{1/3},
	\label{eq:gamma}
\end{eqnarray}
where $k_{\rm B}$ is
Boltzmann's constant, $T_{c,8}\equiv T_c/10^8\ {\rm K}$, $Z_c$ is the
charge of the crust nuclei, and
$a_c = (3/4\pi n_{i,c})^{1/3}$ is the average ion spacing
with $n_{i,c}$ the ion number density in the crust. This transition occurs
at $\Gamma\approx173$ (Farouki \& Hamaguchi 1993 and references
therein), implying a density at the top of the crust,
\begin{eqnarray}
\rho_c &=& 1.1\times10^9\textrm{ g cm}^{-3}
                \left(\frac{T_{c,8}}{3}\right)^3
                \nonumber
                \\
                &&\times
                \left(\frac{A_c}{64}\right)
                \left(\frac{30}{Z_c}\right)^6
                \left(\frac{\Gamma}{173}\right)^3.
                \label{eq:density}
\end{eqnarray}
Since material in the crust is simply ocean material that was advected
there by accretion, we assume that the ocean and crust have the same
composition.

\begin{figure}
\plotone{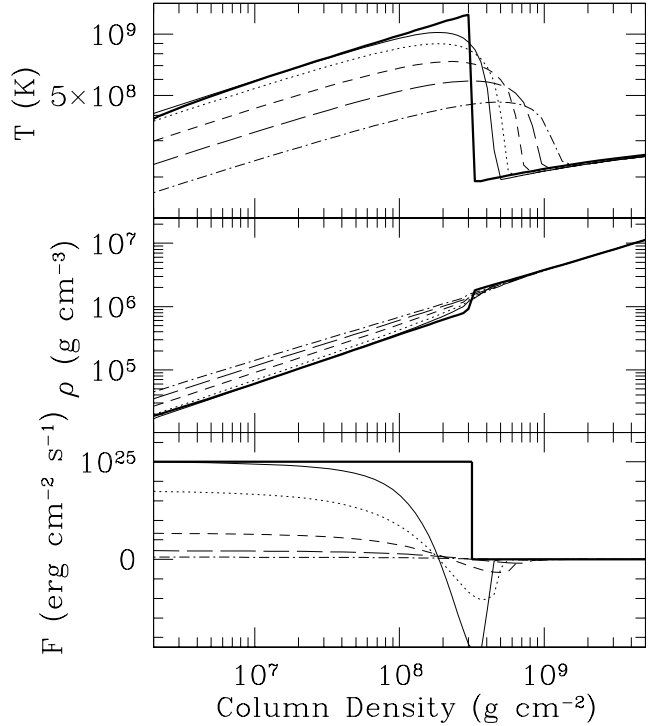}
\caption{Temperature, density, and flux for a He-rich bursting NS
envelope (Model 1). The initial profile is a flux of
$10^{25}\textrm{ erg s}^{-1}\textrm{ cm}^{-2}$
above a column of $3\times10^{8}\textrm{ g cm}^{-2}$, with a flux of
$10^{21}\textrm{ erg s}^{-1}\textrm{ cm}^{-2}$ below ({\it thick
solid lines}). The curves show
time steps of $0.1\ {\rm s}$ ({\it solid}), $0.3\ {\rm s}$ ({\it dotted}),
$1.0\ {\rm s}$ ({\it short-dashed}), $3.0\ {\rm s}$ ({\it long-dashed}),
and $10.0\ {\rm s}$ ({\it dot-dashed}). The majority of the time dependent
change occurs in the outer hot bursting layer where the local thermal time is
short. These profiles extend down into the crust
(which begins at a column of $5.3\times10^{12}\textrm{ g cm}^{-2}$),
but here we just focus on the region where the main changes occur.}
\label{fig:atmos} 
\end{figure}

  We next construct our envelope models using the parameters summarized
above and follow their evolution forward in time.
Cooling is described by the heat diffusion equation with no source terms,
\begin{eqnarray} 
        c_p\frac{\partial T}{\partial t}
		= \frac{\partial F}{\partial y},
	\label{eq:heatdiff}
\end{eqnarray}
where $c_p$ is the heat capacity at constant pressure, and the radiative
flux is
\begin{eqnarray}
        F = \frac{4acT^3}{3\kappa}\frac{\partial T}{\partial y},
        \label{eq:fluxdiff}
\end{eqnarray}
where $a$ is the radiation constant and $\kappa$ is the opacity.
The opacity is set using electron-scattering \citep{pac83},
free-free (Clayton 1993 with the Gaunt factor of Schatz et al. 1999),
and conductive opacities (Schatz et al. 1999 using the basic form of
Yakovlev \& Urpin 1980), and we assume that the crust opacity
is set in the same way. We solve for $\rho$ using the analytic equation
of state from \citet{pac83}. The initial temperature profile is found
by integrating equation (\ref{eq:fluxdiff}) with the flux profile described
above. The envelope is then evolved in time according
to equation (\ref{eq:heatdiff}) using finite differencing techniques.
We use a grid that is uniform in $\sinh^{-1}[\log(y/y_{\rm b})]$
\citep{cm04}, so as to properly resolve the temperature jump at the burning
depth. Figure \ref{fig:atmos} shows the initial profile along with the
profiles at time steps from $0.1-10.0\ {\rm s}$ for Model 1.
The majority of the evolution is
concentrated in the outer hot bursting layer with a negligible thermal wave
diffusing into the ocean and crust, due to the
long thermal time at these depths \citep[consistent with the work of][]{woo04}.
These general features are shared by all of the models we consider.


\section{Nonradial Modes During X-ray Burst Cooling}
\label{sec:modes}

  We now calculate the time dependent spectrum of mode frequencies,
focusing on the surface and the crustal interface waves. Even though
the NS is spinning quickly, $\Omega/(2\pi)\approx270-620\ {\rm Hz}$,
the buoyancy of the envelope dominates over Coriolis effects and
determines the radial structure of the modes (BUC96). This is because
$N^2\sim10^{10}\ {\rm s^{-2}}\gg R\Omega\omega/H_c\sim10^{6}\ {\rm s^{-2}}$,
where the internal buoyancy of the envelope is measured by the
Brunt-V\"{a}is\"{a}l\"{a} frequency
\begin{eqnarray}
        N^2 = -g\left( \frac{d\log\rho}{dz}
                        -\frac{1}{\Gamma_1}\frac{d\log P}{dz}\right),
\end{eqnarray}
and $\Gamma_1 \equiv (\partial\log P/\partial\log\rho)_{s}$ is the
adiabatic exponent. Furthermore, since the modes are squeezed into a thin
outer layer of thickness $\approx H_c\ll R$, they have predominantly
transverse amplitudes. Together these properties justify accounting for
the effects of rotation by using the ``traditional approximation'' (BUC96).
This simplification allows the angular part of
the perturbation equations to be separated from the radial part, and solved
as an eigenvalue equation for $\lambda$ (the ``effective wavenumber''), where
$\lambda$ is a function of $\Omega$ and $\omega$. For this reason, we
focus on the radial equations first, and then introduce the angular and
spin dependent parts afterwards in \S \ref{sec:rotation}.

\subsection{Radial Mode Equations}
\label{sec:radial}

  Solving for the radial eigenfunctions requires considering the governing
mode equations both in the bursting layer and ocean, and also in the NS crust.
The bursting layer and ocean have no shear modulus, so adiabatic perturbations
in this region are described
by the nonradial oscillations equations for an inviscid fluid in
hydrostatic balance and plane-parallel geometry \citep[][hereafter BC95]{bc95}
\begin{eqnarray}
        \frac{d\xi_z}{dz}
        - \frac{\xi_z}{\Gamma_1H}
        = 
        \left( \frac{gHk^2}{\omega^2}- \frac{1}{\Gamma_1 } \right)
        \frac{\delta P}{P},
        \label{eq:mode1}
        \\
        \frac{d}{dz} \frac{\delta P}{P}
        - \left( 1 - \frac{1}{\Gamma_1}  \right) \frac{1}{H} \frac{\delta P}{P}
        = \left( \frac{\omega^2}{g} - \frac{N^2}{g} \right) \frac{\xi_z}{H},
        \label{eq:mode2}
\end{eqnarray}
where $\xi_z$ is radial displacement and $\delta P/P$ is the Eulerian
pressure perturbations.
The Eulerian perturbations are related to the Lagrangian perturbations
by $\Delta P = \delta P - \xi_z\rho g$.
When calculating {\it g}-modes (including the shallow surface wave), equations
(\ref{eq:mode1}) and (\ref{eq:mode2}) are sufficient since these modes are
excluded from the NS crust (BC95). The only exception is the crustal
interface mode (PB05), where it is crucial to solve the mode equations
with a nonzero shear modulus, $\mu$, in the NS crust (BC95). For a classical
one-component plasma the shear modulus is \citep{str91}
\begin{eqnarray}
	\mu = \frac{0.1194}{1+0.595(173/\Gamma)^2}
		\frac{n_{i,c}(Z_ce)^2}{a_c},
\end{eqnarray}
where $\Gamma$ is given by equation (\ref{eq:gamma}). Since
the pressure in the crust is dominated by degenerate, relativistic
electrons we rewrite $\mu$ as
\begin{eqnarray}
	\frac{\mu}{P}
	= \frac{1.4\times10^{-2}}{1+0.595(173/\Gamma)^2}
	\left( \frac{Z_c}{30} \right)^{2/3}.
\end{eqnarray}
In the crust $\mu/P$ is fairly independent of temperature (except for
a small dependence in the factor of $\Gamma$ in the denominator),
so we substitute $\mu_0\equiv\mu/P$ and assume that $\mu_0$ is
constant with depth (PB05).

\subsection{Rotational Effects}
\label{sec:rotation}

  Before the mode equations can be integrated, we must
consistently set $\lambda$ with respect to $\omega$ and $\Omega$.
The dispersion relation that relates these three variables is described
by BUC96 (from the work of Longuet-Higgins 1968), and results in a large
variety of angular eigenfunctions for a given radial structure. \citet{hey04}
showed that very few of these (just the $m=1$ buoyant {\it r}-modes)
match the properties of burst oscillations. We summarize these results here,
both for completeness and because we consider the presence of additional modes
in \S \ref{sec:othermodes}.

\citet{ma04} have extended the traditional approximation to include
general relativity, which introduces
corrections to the co-rotating mode frequencies, $\omega$, due to
frame-dragging and red-shifting. Each of these effects
contribute differently depending on the angular eigenfunction under
consideration, and combined they can decrease
$\omega$ by as much as 20\%. Since the buoyant {\it r}-modes
have frequencies independent of spin
(as we describe below), we expect frame-dragging to be negligible and
red-shifting to be the main change to the normalization of $\omega$
\citep[as is found for the Kelvin mode by][]{ma04}.
From equations (\ref{eq:surface2}) and (\ref{eq:crust2}) we see that such
corrections are degenerate with slight changes in the radius and
temperature of the NS, and therefore we do not include them.

\begin{figure}
\plotone{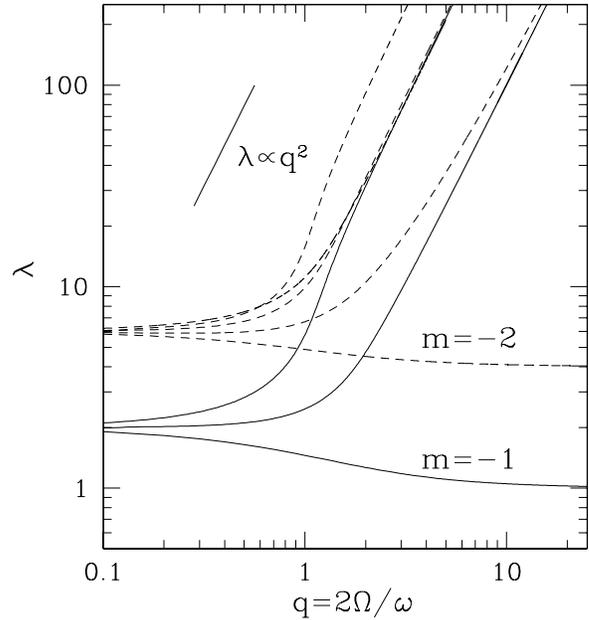}
\caption{Effective wavenumber $\lambda$ as a
function of the spin parameter $q=2\Omega/\omega$ for the rotationally modified
{\it g}-modes ($m\neq-l$) and Kelvin modes ($m=-l$). Solid lines denote
modes for which $l=1$ in the non-rotating limit with $m=-1$, $0$, and $1$
({\it from bottom to top}). The dashed lines
denote modes for which $l=2$ in the nonrotating limit with $m=-2$, $-1$,
$0$, $1$, and $2$ ({\it from bottom to top}). For $q\gg1$, $\lambda\propto q^2$,
except if $m=-l$, which asymptotes to $\lambda=m^2$
(the Kelvin modes). Unfortunately, none of these dispersion relations are
consistent with the burst oscillations because they either have frequency
shifts much too large (in the case of the rotationally modified
{\it g}-modes) or shifts with the wrong sign (in the case of the Kelvin modes).}
\label{fig:gmodes}
\end{figure}

  There are two sets of solutions for $\lambda$, which are both typically
presented in the literature as a function of the ``spin parameter,''
$q=2\Omega/\omega$ (so that for a quickly spinning star $q\gg1$) The
first set is comprised of the rotationally modified {\it g}-modes (those with
$m\neq-l$) and the Kelvin modes ($m=-l$),
which are shown in Figure \ref{fig:gmodes}. In the slowly rotating limit,
$q\ll1$, we find $\lambda\approx l(l+1)\approx2$ or $6$ (for $l=1$ or
$2$, respectively), but as the spin increases $\lambda$ splits depending
on its $m$ value. It then asymptotes to scaling as $\lambda\propto q^2$
(in the case of the rotationally modified {\it g}-modes) or as $\lambda=m^2$
(in the case of the Kelvin modes).

  We present these solutions to emphasize
that none of these modes are consistent with burst oscillations.
The rotationally modified {\it g}-modes all have $\lambda\gg1$ when
the star is rotating quickly. This implies large mode frequencies
(see eqs. [\ref{eq:surface2}] and [\ref{eq:crust2}]), and
frequency drifts larger than observed. (In \S \ref{sec:othermodes} we explore
whether these modes exist in burst cooling light curves, but are not seen
{\it because of their large frequency drifts}.)
The Kelvin modes are inconsistent because they are all prograde
($m<0$), so that as the surface cools an observer would see a
{\it decreasing} frequency.

\begin{figure}
\plotone{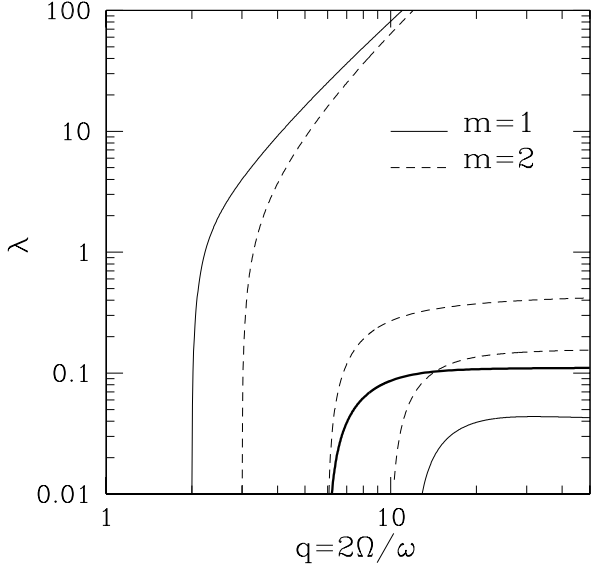}
\caption{Effective wavenumber $\lambda$ as a function of
the spin parameter $q=2\Omega/\omega$ for the {\it r}-modes. Solid lines
denote modes for which $m=1$, with values of $l$ (identified in the slowly
rotating limit using eq. [\ref{eq:rmode}]) of $1$, $2$, and $3$ ({\it from left to
right}). Dashed lines denote modes for which $m=2$, with
$l=1$, $2$, and $3$ ({\it from left to right}). We favor the $l=2$, $m=1$
mode ({\it thick line}) as explaining the burst oscillations, which
has $\lambda\approx1/9\approx0.11$ in the quickly rotating limit.}
\label{fig:rmodes}
\end{figure}

  The other set of solutions are a group of modes unique to the
case of a rotating star, the {\it r}-modes, which we plot in
Figure \ref{fig:rmodes}. These occur as zero frequency solutions
for a non-rotating star, corresponding to incompressible toroidal
displacements on the stellar surface. When the star is rotating, Coriolis effects
turn these solutions into normal modes of oscillation. In the slowly rotating
limit the NS spin and mode frequency are directly proportional,
\be
	\omega_r=\frac{2m\Omega}{l(l+1)},
	\label{eq:rmode}
\ee
\citep{sai82,ls86}, which are the well-studied inertial {\it r}-modes.
More relevant for our work here is the case when $q\gg1$, which are the
buoyant {\it r}-modes.  For $q\gg1$ the
{\it r}-modes with $m=l$ exhibit $\lambda\propto q^2$, similar
to the rotationally modified {\it g}-modes from Figure \ref{fig:gmodes},
and thus also give frequency shifts too large for burst oscillations.
More promising are the {\it r}-modes with $m<l$, which have $\lambda\lesssim1$
and therefore small frequency shifts. Furthermore, since all {\it r}-modes
have $m>0$, they are traveling retrograde with respect to the NS spin, and
provide {\it increasing} shifts as the NS cools. We favor the $l=2$, $m=1$
mode (denoted with a thick line in Figure \ref{fig:rmodes}) over any of the other
$\lambda\ll1$ {\it r}-modes because: (1) $m=1$ is implied from the
observations and this is the lowest order $m=1$ mode that gives frequencies
and shifts consistent with burst oscillations and (2) any higher order
$\lambda\ll1$ {\it r}-mode has multiple bands of hot and cold regions
at different latitudes, so that it should be more difficult to observe than the
$l=2$, $m=1$ mode. It is rather remarkable that with the number of
rotationally modified modes that exist, this is the only one that fits
all the required properties! For this mode $\lambda\approx1/9\approx0.11$
in the quickly rotating limit, as we used when estimating the mode
frequencies in \S \ref{sec:sum}.

  A further reason for favoring the buoyant {\it r}-modes is that
their latitudinal eigenfunctions may be easier to observe \citep{hey04}. Modes
with $\lambda\gg1$ have their eigenfunction squeezed near the equator within
an angle $\cos\theta<1/q$, where $\theta$ is measured from the pole. On the
other hand, the $m<l$ {\it r}-modes and Kelvin modes have much wider
eigenfunctions that span most of the surface. This means that given a fixed
equatorial perturbation, the pulsed fraction is larger for these latter modes.

\subsection{Eigenfunctions and Frequencies in Rotating Frame}
\label{sec:timedependent}

  Normal modes of oscillation are found by assuming $\Delta P=0$ at
the top boundary, which is set at a depth where the local thermal time is
equal to the mode period ($t_{\rm th}=2\pi/\omega$, where
$t_{\rm th}\equiv c_pyT/F$). This top condition, though not unique,
is fairly robust since little mode energy resides in the low density
upper altitudes (BC95). This is especially true for the two modes we study
here, which have their energy concentrated at their respective
interface. We numerically integrate the mode equations, shooting
for the condition that $\xi_z\approx\xi_x\approx0$
deep within the crust, where $\xi_x$ is the transverse displacement.
This may not be the case in a more realistic calculation of the crust,
but as long as we set
this bottom depth deep enough, $\approx10^{17}\ {\rm g\ cm^{-2}}$,
we recover the asymptotic solutions of PB05. When we change
from integrating the non-viscous mode equations to those appropriate
in the crust, we must properly set boundary conditions (PB05).

\begin{figure*}
\epsscale{1.0}
\plotone{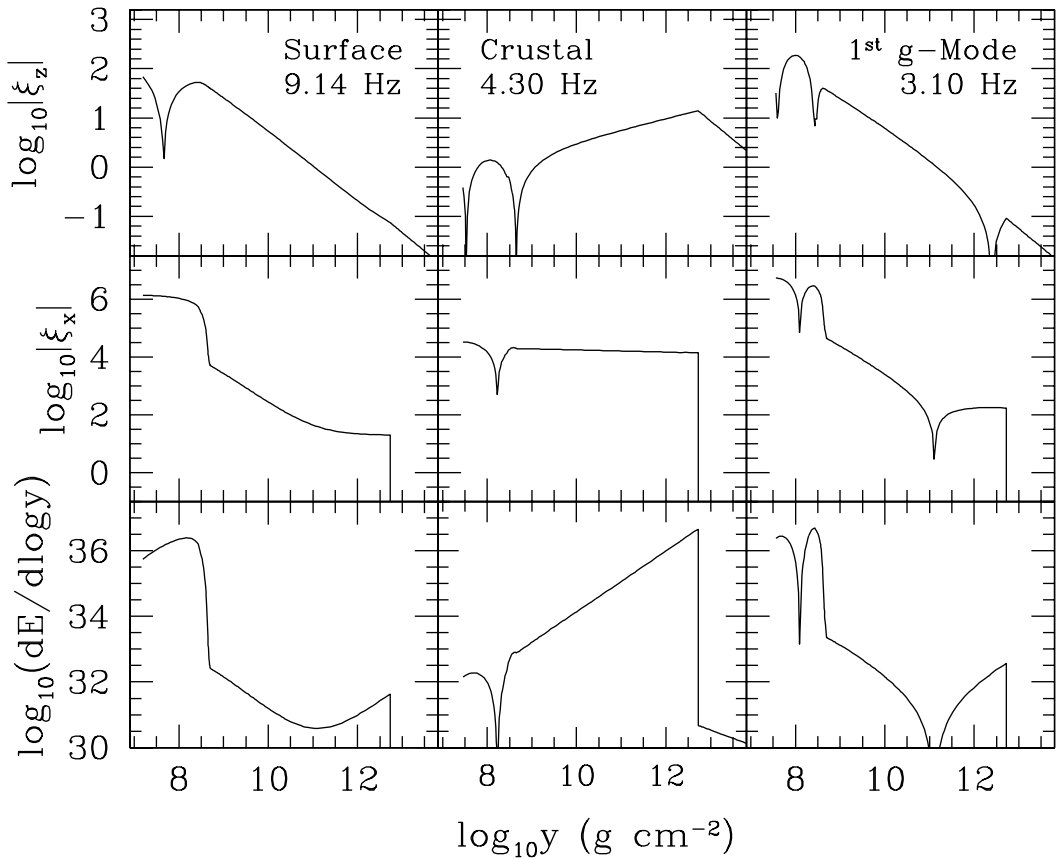}
\caption{Displacement eigenfunctions (given in cm) and energy density
(measured in ergs and given by eq. [\ref{eq:energy}]) of the first three low
frequency solutions. These are from Model 1 after $0.1\ {\rm seconds}$
of cooling and all with $\lambda=0.11$.
The frequencies are in the rotating frame on the NS surface, $\omega$.
From high to low frequency ({\it left to right}), these are the shallow surface wave,
the crustal interface wave, and the first {\it g}-mode. All lower frequency
solutions
are also {\it g}-modes. Each lower frequency mode contains an additional
node, which are shown by the cusps because we plot the absolute value
of the displacements. The transverse displacement, $\xi_x$, always
has a node at the crust at a column of $5.3\times10^{12}\ {\rm g\ cm^{-2}}$.}
\epsscale{1.0}
\label{fig:modeshapes}
\end{figure*}
  When solving the mode equations we use an effective wave number
$\lambda\approx0.11$, as described in \S \ref{sec:rotation}.
There are many solutions to the mode equations, which form a
complete basis set of functions. We focus on the low frequency solutions
(as opposed to the {\it f}-mode or {\it p}-modes) because they are a better
match to the observed frequencies. These solutions are ordered from high to low
frequency, with each successive mode having an additional node in its
radial eigenfunctions. In Figure \ref{fig:modeshapes} we plot the radial and
transverse displacement eigenfunctions ($\xi_z$ and $\xi_x$, respectively,
given in cm) for the first three low frequency solutions from Model 1 after
$0.1\ {\rm s}$ of cooling. In addition we plot the energy per logarithm column
density,
\begin{eqnarray}
        \frac{dE}{d\log y}
        = \frac{1}{2}4\pi R^2 \omega^2\xi^2 y,
        \label{eq:energy}
\end{eqnarray}
where $\xi^2=|\xi_x|^2+|\xi_z|^2$ is the total displacement. This indicates
where the kinetic energy of the mode is
concentrated. We normalize the total integrated energy of each mode
to be $5\times10^{36}\ {\rm ergs}$, $10^{-3}$ of the total energy
released by unstable nuclear burning
\citep{bil98}. At $9.14\ {\rm Hz}$ (where this is the frequency in the rotating
frame on the NS surface, $\omega$) we find a mode with a single node in its
eigenfunctions (shown as a cusp in $\xi_z$ and less apparent in $\xi_x$
because there is always a node at the crust at a column of
$y_c=5.3\times10^{12}\ {\rm g\ cm^{-2}}$). Since the energy is
concentrated in the bursting layer this is identified as the shallow surface
wave. The next mode at $4.30\ {\rm Hz}$ has an additional node and therefore
a different parity than the surface wave. Its energy is distributed very
differently, largely concentrated at the bottom of the ocean, so it is
identified as the crustal interface mode. All modes with lower frequencies are
{\it g}-modes, and we show the first {\it g}-mode (at
$3.10\ {\rm Hz}$) with three nodes.\footnote{These {\it g}-modes should not
be confused with the {\it rotationally modified}
{\it g}-modes from \S \ref{sec:rotation}. Here we are only considering the
radial eigenfunction. In the previous case we were discussing the angular
eigenfunctions, which can be applied to any radial eigenfunction.}
These modes are trapped in either the bursting layer or
ocean and are identified by a relatively constant energy distribution in
each region (the example shown is the prior case).

  Since both the surface wave and the first {\it g}-mode are concentrated in the
bursting layer, they each exhibit frequency drifts as the bursting layer cools,
making them both attractive as being burst oscillations. We therefore consider
the radiative damping time for each mode to narrow down the choice of
radial eigenfunction. The rate of energy loss averaged over one oscillation
cycle from an adiabatic perturbation is estimated from the ``work integral''
\citep{cox80,unn89}, which, when there is no energy source, is
\be
	\frac{dE}{dt} = \frac{\omega}{2\pi}\oint dt
		\int\frac{\Gamma_3-1}{\Gamma_1}\frac{\Delta P}{P}
		\Delta\left(\frac{dF}{dy}\right)4\pi R^2dy,
	\label{eq:workintegral}
\ee
where $\Gamma_3-1 \equiv (\partial\log T/\partial\log\rho)_{s}$. The integral
is negative when a mode is damped and positive when a mode is excited.
From this and the total mode energy (integrating
eq. [\ref{eq:energy}]) we can estimate the damping $e$-folding time
of a mode's amplitude,
\be
	t_{\rm damp} = \left(\int\frac{dE}{d\log y}d\log y\right)
			\left| \frac{dE}{dt}\right|^{-1},
\ee
where the absolute value has been included to make this timescale positive
(since all modes we consider are damped).

\begin{figure}
\plotone{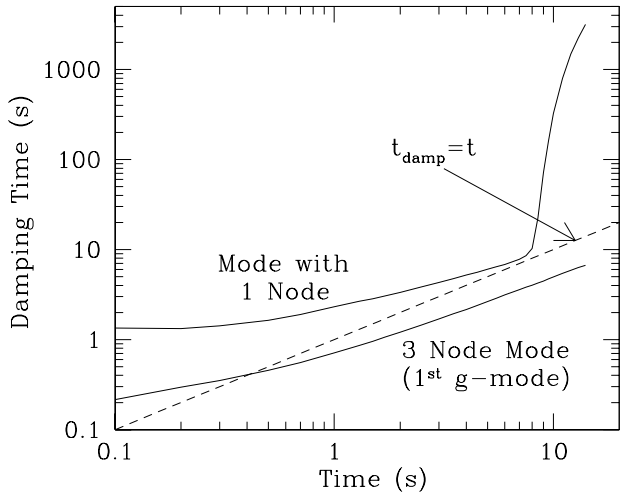}
\caption{Damping time, $t_{\rm damp}$, as a function of the time since
the burst peak, for the mode with a single
node (which is initially a surface wave) and the mode with three nodes
(the first {\it g}-mode), both from Model 1. The dashed line shows were
the damping time is equal to the time since burst peak. After only
$\approx0.5\ {\rm s}$ the {\it g}-mode is exponentially damped on a timescale
shorter than $0.5\ {\rm s}$, so that it will not be observable.}
\label{fig:damping}
\end{figure}

  Figure \ref{fig:damping} shows $t_{\rm damp}$ for both the surface
wave and the first {\it g}-mode as functions of time since the burst peak.
To accurately calculate the integral from equation
(\ref{eq:workintegral}) we must drop derivatives of $\Delta P/P$ near the
surface because our imposed boundary condition of $\Delta P=0$ results in
unphysical local mode excitation \citep[for a description of another way
to set this boundary condition see][]{gw99}. The {\it g}-mode
is damped on a timescale an order of magnitude faster than the surface
wave. Though both the first {\it g}-mode and the surface wave have energies
concentrated in the bursting layer (see far left and right panels at the bottom
of Figure \ref{fig:modeshapes}), each mode's energy {\it distribution} is
different within this region.
Radiative damping predominantly takes place at the top of the atmosphere
\citep{pb04}, so that a mode with more of its energy distributed there is
more damped. It is somewhat difficult to see because of the large dynamic
range we show, but the energy of the {\it g}-mode is approximately constant
in the bursting layer, while the shallow surface wave's energy varies by almost
an order of magnitude in the bursting layer (compare the bottom panels on the
far left and right of Figure \ref{fig:modeshapes}).
The ratio of damping timescales is therefore approximately
the ratio of energies between the top and bottom of the bursting layer.
After $\approx0.5\ {\rm s}$ the {\it g}-mode has a damping time shorter
than the time for the burst to cool, so that the amplitude
of this mode is exponentially damped. Higher-order {\it g}-modes have different
damping times depending on where they are trapped. Those trapped in
the bursting layer (like the example shown) all have similar, short damping
times. The {\it g}-modes that are trapped in the ocean have much longer
damping times ($t_{\rm damp}\approx5-10\ {\rm s}$), but suffer from very
small surface amplitudes.
The crustal wave has an even longer damping time
($t_{\rm damp}\gtrsim10^3\ {\rm s}$), but also has the problem of a small
surface amplitude. We therefore favor the surface wave as the radial
eigenfunction for the burst oscillations because of its combined attributes
of having a long damping time coupled with a large surface amplitude from
a small input of energy.

\begin{figure}
\plotone{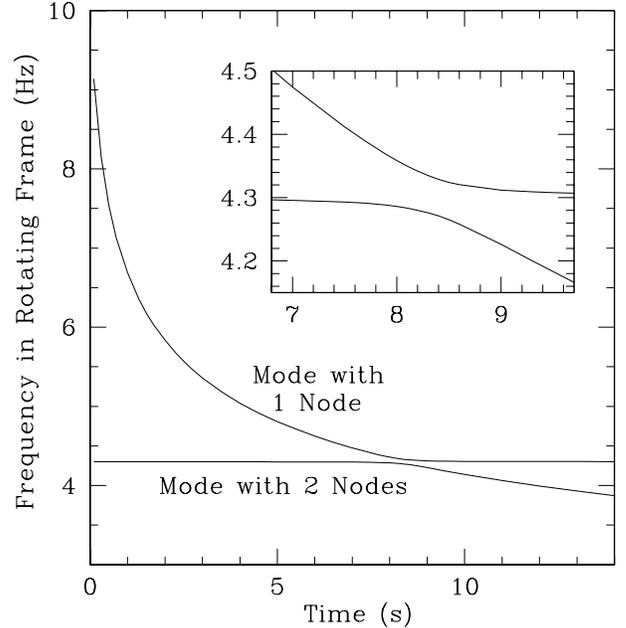}
\caption{The frequency evolution of the surface wave and crustal
interface wave using Model 1. The upper curve follows the mode with
one node in its eigenfunctions, while the lower curve follows the mode with
two. The qualitative shapes of the eigenfunctions vary
considerably along each curve. Initially, the higher frequency mode
corresponds to a shallow
surface wave, while the lower frequency mode is a crustal interface wave.
As the lines move to the right toward later times, the surface wave is sensitive
to the cooling surface layers and decreases, while the crustal wave
does not evolve. Eventually the two modes meet at an avoided crossing,
which we display in an inset panel. Past the avoided
crossing, although the parity has not changed, the higher frequency
mode is now like a crustal wave, while the low frequency mode is a surface
wave.}
\label{fig:nonrotating}
\end{figure}

  We next consider how the mode frequencies evolve with time since burst
peak. In Figure \ref{fig:nonrotating} we plot the first two eigenfrequencies
as a function of time for Model 1. The mode parity does not change
along each continuous
line, with one node present for the upper line and two nodes for the
lower line. Initially the upper frequency is the surface wave and the lower
is the crustal interface wave, consistent with equations (\ref{eq:surface2}) and
(\ref{eq:crust2}). The surface wave's frequency decreases as the
bursting layer cools, eventually running into the crustal mode. This
creates an avoided crossing between the two modes (which we
highlight in the inset panel of Figure \ref{fig:nonrotating}).
The eigenfunctions therefore change considerably
along each line from being characteristic of a surface wave to a crustal
interface wave (and vice versa). As long as the evolution of the envelope
is sufficiently slow, once excited a mode will most likely evolve along
one of these branches and not be able to skip to the other because of
differences in parity. We do not show this rigorously, but think this is
a reasonable conclusion by considering analogies with perturbation
theory problems in non-relativistic quantum mechanics. In cases where
the background changes quickly, on a timescale shorter than
$\sim2\pi/\omega$, mixing is highly possible as determined by taking
inner products between initial and final states on either side of the
avoided crossing. On the other hand, in the case we consider here the
background changes very slowly, as is supported by the fact that
$d(1/\omega)/dt\lesssim10^{-2}\ll1$ (which we calculated to check
our conclusions). This implies that the change can be
treated as adiabatic, and the modes are not expected to mix at the crossing.

\begin{figure}
\plotone{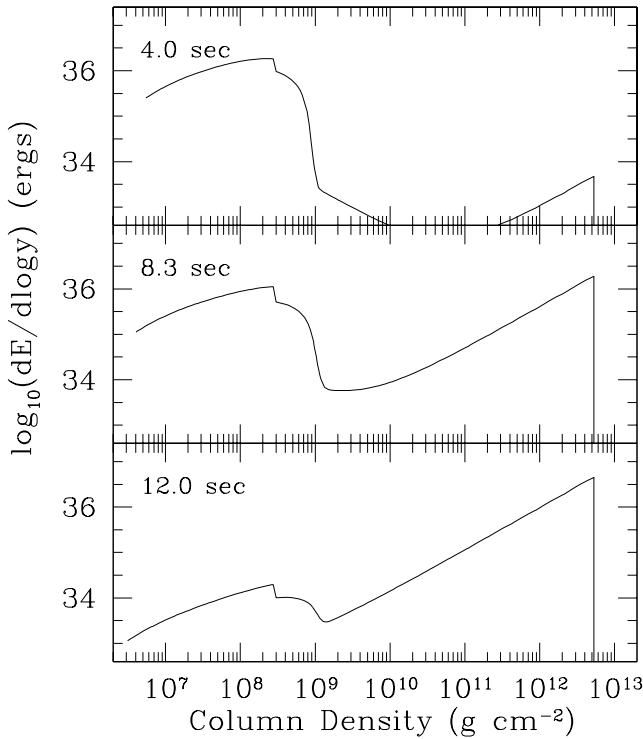}
\caption{Energy of the single-node mode per unit logarithmic pressure at
different time steps, from Model 1. Each mode is normalized to a total energy
of $5\times10^{36}\ {\rm ergs}$, $10^{-3}$ of the total
energy released in an X-ray burst. At early times the mode looks like
a shallow surface wave, with its energy concentrated near in the
bursting layer, but as it passes through the avoided crossing (see
Figure \ref{fig:nonrotating}) its energy moves continually deeper and
becomes more like a crustal interface wave. The sharp feature at a column of
$3\times10^8\ {\rm g\ cm^{-2}}$ is due to the density discontinuity at
the burning depth. Below a column of
$5.3\times10^{12}\ {\rm g\ cm^{-2}}$ the crust begins and the mode's
energy drops off quickly.}
\label{fig:energy}
\end{figure}

 In Figure \ref{fig:energy} we plot the energy of the single-node wave for a
sample of time steps from Model 1. We normalize
the total integrated energy of each mode to be $5\times10^{36}\ {\rm ergs}$.
Initially, at $4.0\ {\rm s}$, the mode's energy is
concentrated in the recently burned portion of the upper
envelope. As time progresses the energy goes deeper into
the star, so that by $12.0\ {\rm s}$, when the burst
is nearly over, the energy is concentrated at the bottom of the
ocean, just as expected for a crustal mode.

\subsection{Observed Frequencies}

\begin{figure}
\plotone{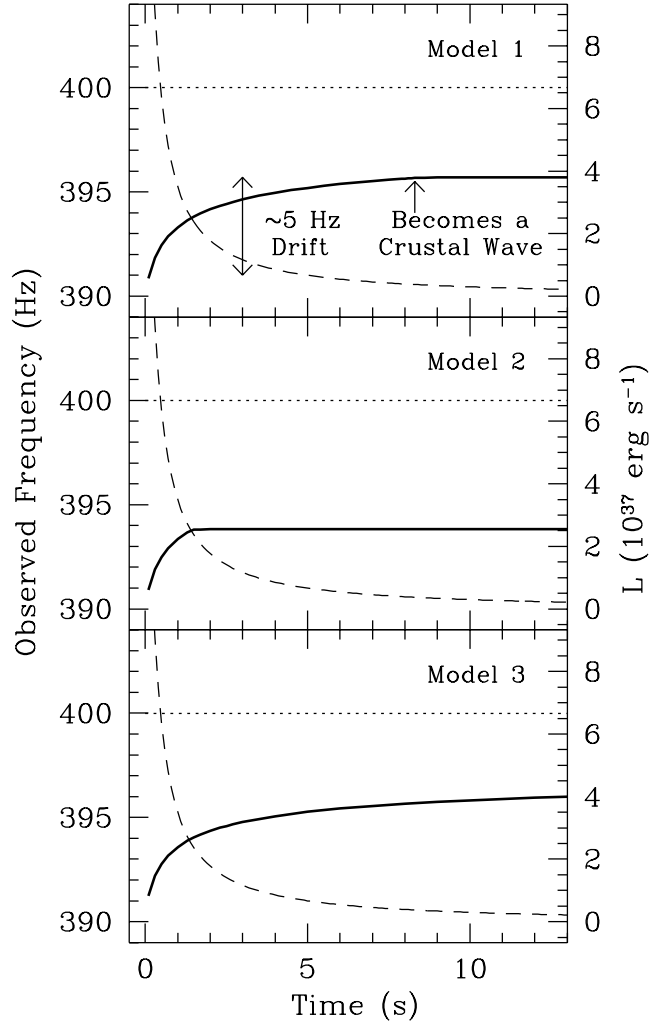}
\caption{The observed frequency evolution of the single-node mode on
Models 1, 2, and 3 ({\it solid curves}).
We use a NS spin of $400\ {\rm Hz}$ ({\it dotted line}). For Model 1, this
shows that while the mode is a surface wave it evolves upward in frequency,
but once
it becomes a crustal wave it is very stable at a frequency of
$\omega_c/(2\pi)\approx4\ {\rm Hz}$ below the NS spin.
In Model 2 the frequency shift ends very early
and is only $\approx3\ {\rm Hz}$ because the crust is much hotter
due to a larger crustal flux. On the other hand, in Model 3 we find that
an avoided crossing between the surface wave and crustal wave does
not occur while the luminosity is at an observable level, so that the frequency
never becomes flat. Observations of a NS with oscillations like this
would show more variance in their asymptotic frequency in
comparison to the other models we consider. The dashed line is
the surface luminosity as a function of time, assuming a radius
of $10\ {\rm km}$.}
\label{fig:rotating}
\end{figure}

  We now focus on the evolving mode with a single node because it is initially
a surface wave, which as discussed in \S \ref{sec:timedependent} is the mode
most least likely to be damped and also have an appreciable amplitude.
In Figure \ref{fig:rotating}
we use equation (\ref{eq:obs}) to plot the frequencies seen by an inertial
observer ({\it solid lines}) for a NS spinning at $400\ {\rm Hz}$
({\it dotted line}) for each of our three models.
We also show the light curves with a dashed line. In the top panel
we plot the results from Model 1.
This shows that the NS spin is $\approx4\ {\rm Hz}$ above the burst
oscillation, just the frequency of the crustal mode.
In the next two panels we plot the frequency evolution of our other
two models, along with their respective light curves. In Model 2, the flux
coming through the crust is a factor of 3 higher, providing a significantly
hotter crust. This causes the crystallization to happen at much larger
depths, resulting in a larger crustal wave frequency. The mode crossing
therefore happens earlier, which can be seen by the smaller frequency
drift, as expected from the analytic
considerations in \S \ref{sec:sum}. In Model 3, there is a slightly larger
discontinuity at the burning depth due to the heavy ashes from
rp-process burning as expected in H/He mixed bursts. This makes the
evolution of the surface wave much slower, so that in this case it does
not transition into a crustal interface wave, even after $14\ {\rm seconds}$ of
cooling. This shows that an avoided mode crossing is not a robust property
exhibited by every model. Observations of burst oscillations from such a NS
would show more variance in its asymptotic frequency because of this
feature. On the other hand, since Model 3 represents an rp-process
burst and such bursts do not exhibit burst oscillations, this has not
been observed. It is not clear from our models why the He models that
show mode crossings are the ones observed, while our
H/He models that don't show crossings are not, especially since
frequency drifts like that shown in the bottom panel
would be observable if they existed. We discuss these issues
further when we conclude in \S \ref{sec:concl}.

  As noted by \citet{hey04}, since $\lambda\approx0.11$ as
long as $q\gtrsim10$, any spin NS will exhibit similar frequency evolution
as long as $\Omega/(2\pi)\gtrsim5\omega_s/(2\pi)\approx50\ {\rm Hz}$. This
makes it easy to compare models with different spins, and constrains the
explanation of burst oscillations as modes
since it requires that the frequency shifts be spin independent. This is
counter to the studies of \citet{cb00}, which predict that the spin and
drift size should be correlated.
Although EXO $0748-676$ is near this critical frequency (see Table 1),
its drift (assuming that it is not yet a crustal interface wave)
is still not heavily affected. We find a $\approx4\ {\rm Hz}$ drift
instead of a $5\ {\rm Hz}$ drift at higher rotation rates, so we still
predict a spin frequency of $\approx49\ {\rm Hz}$ for this object.

\subsection{Surface Amplitude Evolution}
\label{sec:ampl}

  As the surface wave evolves into a crustal wave, we expect the surface
amplitude to decrease dramatically since the crustal wave's energy is mostly
at the bottom of the ocean. We estimate $\delta P/P$ at our
``surface'' where the modes are no longer adiabatic, $t_{\rm th}=2\pi/\omega$,
denoted $(\delta P/P)_s$, which provides a general idea for how the amplitude
is changing. BC95 show that the Eulerian perturbation $\delta P/P$
is constant in the non-adiabatic region,
so that we only need to calculate it at our top boundary to estimate its value
at the photosphere. Approximately $5\times10^{39}\ {\rm ergs}$ of energy is
released in an X-ray burst, so we assume some small fraction
of this is able to power the mode. Since we do not know how much energy
actually leaks into the mode, we simply study values which result
in reasonably sized perturbations. A key conclusion is that only a small fraction
of the available energy is needed, $\sim10^{-3}$, to get perturbations of order
unity. We assume that this energy is conserved throughout the evolution of the
mode, an approximation that is justified because the damping time of the mode
is always longer than the time that the burst has been cooling.

\begin{figure}
\plotone{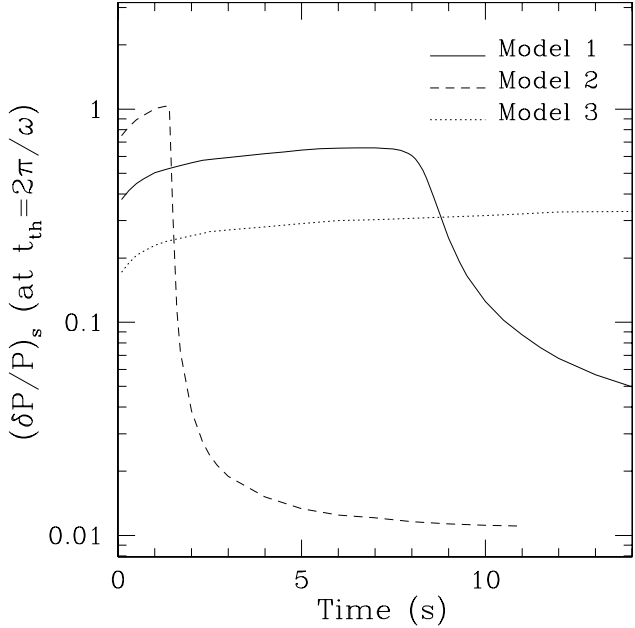}
\caption{Evolution of the surface pressure perturbation for the single-node
mode from each of our
models. In each case we assume a certain fraction of the X-ray burst's
$5\times10^{39}\ {\rm ergs}$ of energy has gone into the modes, and is then
conserved in the mode as the envelope cools. The fractions we use
are $10^{-3}$ (model 1; {\it solid line}), $4\times10^{-3}$ (model 2; {\it dashed line}),
and $2\times10^{-4}$ (model 3; {\it dotted line}). This demonstrates that appreciable
amplitudes can be expected, even if only a small fraction of the burst's
energy goes into the modes. This also highlights how the amplitude quickly
dies off once the shallow surface wave turns into a crustal interface mode.}
\label{fig:ampl}
\end{figure}
  In Figure \ref{fig:ampl} we plot the surface perturbation as a function of
time for all three models. In both Models 1 and 2 the perturbation drops
off dramatically once the surface wave turns into a crustal interface wave,
especially in the case of Model 2, which has a thicker ocean. If the top of the
crust is too deep the resulting
surface amplitude will be too small to explain burst oscillations. The total
displacement is about constant down to the crust with
$\xi\approx\xi_x\approx\xi_z R/h$ since the modes are nearly
incompressible. From our top boundary condition of $\Delta P=0$ we find
$(\xi_z/h)_s=(\delta P/P)_s$, so that together $\xi\sim(\delta P/P)_sR$.
Using the scalings from equation (\ref{eq:energy}), along with the crustal mode
frequency from equation (\ref{eq:crust2}) and the density of crystallization
from equation (\ref{eq:density}), the total crustal wave energy at late times
is approximately
\be
	E_{{\rm total},c} \sim 5\times10^{36}\ {\rm ergs}
		\lp\frac{64}{A_c}\rp
		\lp\frac{30}{Z_c}\rp^{20/3}
		\lp\frac{R}{10\ {\rm km}}\rp^2
		\nonumber
		\\
		\times\lp\frac{\lambda}{0.11}\rp
		\left[\frac{(\delta P/P)_s}{0.1}\right]^2
		\lp\frac{T_c}{3\times10^8\ {\rm K}}\rp^5,
\ee
where the prefactor has been set to match the numerical results. The predominant
effect is that the surface amplitude is very sensitive to the depth of the crust
so that $(\delta P/P)_s\propto Z_c^{20/3}T_c^{-5/2}$ for fixed energy.
In Model 3 the switch to a crustal interface wave never occurs, so that the
amplitude always stays large. An interesting feature of all the models is that
at early times the amplitude actually increases. This is because as the frequency
initially decreases the displacements must increase to maintain the same
energy. Suggestively, the time evolutions of the amplitudes, especially
for Model 1, are similar to the amplitudes measured by \citet{moc02}.


\subsection{Magnetic Field Limits}
\label{sec:magfield}

  The lack of persistent pulsations from the non-pulsar burst oscillation sources
implies a smaller magnetic field for these objects. To prevent channeled
accretion the surface dipole field must be less than \citep{fkr92}
\be
	B_{\rm acc}=5.1\times10^7\ {\rm G}\lp\frac{M}{1.4M_\odot}\rp^{1/4}
		\lp\frac{10\ {\rm km}}{R}\rp^{5/4}
		\nonumber
		\\
		\times\lp\frac{\dot{M}}{10^{-9}\ M_\odot\ {\rm yr^{-1}}}\rp^{1/2}.
\ee
Both the shallow surface wave and the crustal interface wave exhibit large
shears in their transverse displacements. Following BC95, the maximum
magnetic field before the waves would be dynamically affected is
$B^2\approx8\pi H^2\rho\omega^2$, where $H$ is chosen as appropriate for
each mode. For the surface wave this gives a limit
(using eq. [\ref{eq:surface2}]),
\be
	B_s=8.8\times10^7\ {\rm G}
	\lp\frac{2Z_b}{A_b}\frac{T_b}{10^9\ {\rm K}}\rp
	\lp\frac{y_b}{3\times10^8\ {\rm g\ cm^{-2}}}\rp^{1/2}
	\nonumber
	\\
	\times\lp\frac{1.4M_\odot}{M}\frac{\lambda}{0.11}\rp^{1/2}
	\lp1-\frac{T_c}{T_b}
		\frac{\mu_b}{\mu_c}\rp^{1/2}.\hspace{0.5cm},
\ee
surprisingly close to $B_{\rm acc}$ considering that these two limits
result from vastly different physical mechanisms. For the crustal interface
mode (using eqs. [\ref{eq:crust2}] and [\ref{eq:density}]),
\be
        B_c=1.2\times10^{10}\ {\rm G}
        \lp\frac{64}{A_c}\rp\lp\frac{30}{Z_c}\rp^{11/3}
	\lp\frac{T_{c,8}}{3}\rp^3
	\nonumber
	\\
        \times\lp\frac{1.4M_\odot}{M}\frac{R}{10\ {\rm km}}\rp
	\lp\frac{\lambda}{0.11}\rp^{1/2},
\ee
where we assume $\Gamma=173$. This is much higher than either of the two
previous limits because the crustal interface wave ``lives'' at such high
pressures. The lack of persistent pulses from the majority of X-ray
bursters is likely because of a weak field in the bursting layer and ocean
\citep{czb01}, making our non-magnetic mode calculation adequate for now.


\section{Comparisons with Observed Drifts}
\label{sec:drifts}

  \citet{bbr98} showed that the core of a transiently accreting NS reaches
a steady-state set by nuclear reactions deep in the crust, so that the
luminosity departing the crust is given by the time-averaged accretion rate
\citep{bro00}. This means that crustal temperatures can be inferred from
an observable property, namely the average persistent luminosity.
Given the discussion from \S \ref{sec:sum}, we expect objects with small
observed
drifts to have larger persistent luminosities.

\begin{figure} 
\plotone{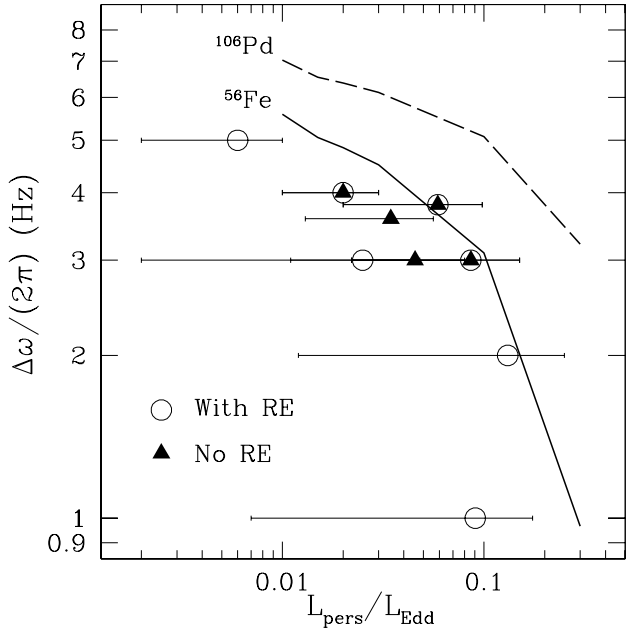}
\caption{The frequency drifts, $\Delta\omega/(2\pi)$, as a function of the
persistent luminosity, $L_{\rm pers}$
(in units of $L_{\rm Edd}=2.5\times10^{38}\ {\rm erg\ s^{-1}}$).
Two different crustal models are plotted (Brown 2005, private communication)
with compositions of $^{56}$Fe ({\it thick solid line}) and $^{106}$Pd
({\it thick dashed line}). We assume that $A_b/Z_b=2$, $\mu_b/\mu_c=1.0$,
and that initially $T_b=10^9\ {\rm K}$. We compare this with the largest
observed frequency drifts of the non-pulsar, bursting NSs (see Table 1).
The observed drifts should
be viewed as {\it lower limits} since observational or
excitation-related effects may result in the initially observed surface
wave having $T_b<10^9\ {\rm K}$. The horizontal bars indicate the range of
luminosities shown by each object, with a point at the average
luminosity. Open circles indicate NSs that show radius expansion bursts
at the same time as burst oscillations, while solid triangles indicate
NSs that do not. 4U $1728-34$, Aql $X-1$, and 4U $1636-536$
have shown both cases.} 
\label{fig:observed}
\end{figure}
  In Figure \ref{fig:observed} we plot
predicted frequency drifts using the difference of equations
(\ref{eq:surface2}) and (\ref{eq:crust2}). We relate $T_c$ to the persistent
luminosity,
$L_{\rm pers}$, using the crustal models of Brown (2005, private communication)
for compositions of $^{56}$Fe ({\it thick solid line}) and $^{106}$Pd ({\it thick dashed
line}). These models are based on the calculational techniques described
in \citet{bro04}. We then
plot the observed luminosities and largest observed frequency drifts from
Table 1. The total range of observed luminosities is shown by the horizontal
bars with a point placed at the average luminosity. Open circles indicate
objects for which burst oscillations are coincident with radius expansion
bursts, while solid triangles indicate those that are not.
We use an Eddington luminosity of
$L_{\rm Edd} =2.5\times10^{38}\ {\rm erg\ s^{-1}}$ so as to be consistent
with the work of \citet{for00}. The observed drifts should be viewed
as lower limits since the surface wave may not be excited
until the bursting layer's temperature is below the
$T_b=10^9\ {\rm K}$ that we use here. Qualitatively, the higher luminosity
NSs show smaller drifts, as expected from our models. Furthermore,
lighter crustal material is favored for explaining the observations. This is
consistent with the cooling timescale of the bursts (see Table 1), which suggest
He-rich bursts that produce iron-like nuclei ( and would not produce
rp-process elements such as
$^{106}$Pd). In general, these objects are not all expected to have the
same crustal composition, and this can be studied by making further
comparisons with detailed modeling and consideration of each system's bursting
and binary properties.


\section{Could Other Angular Modes be Present?}
\label{sec:othermodes}

\begin{figure*}
\epsscale{1.0}
\plottwo{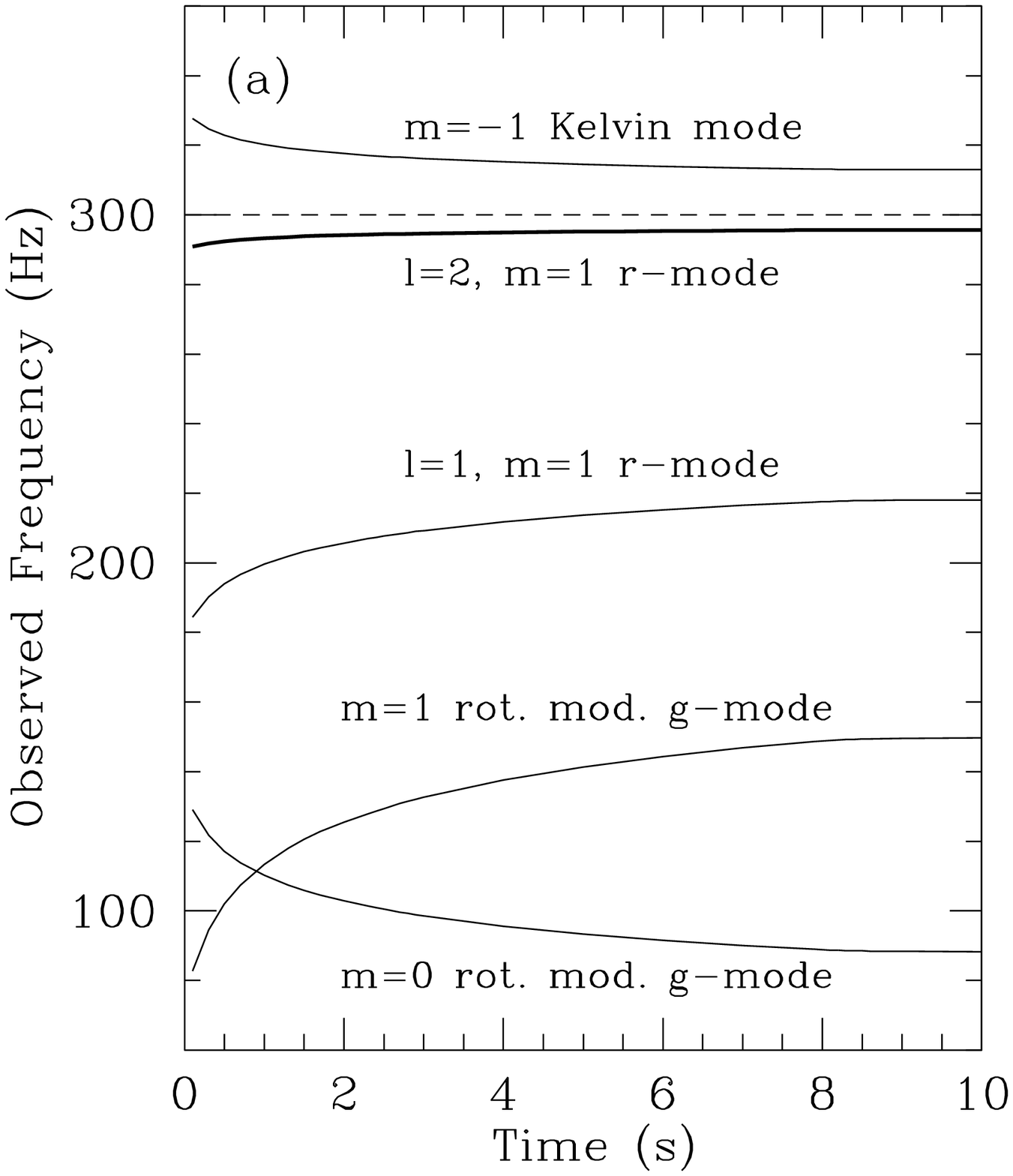}{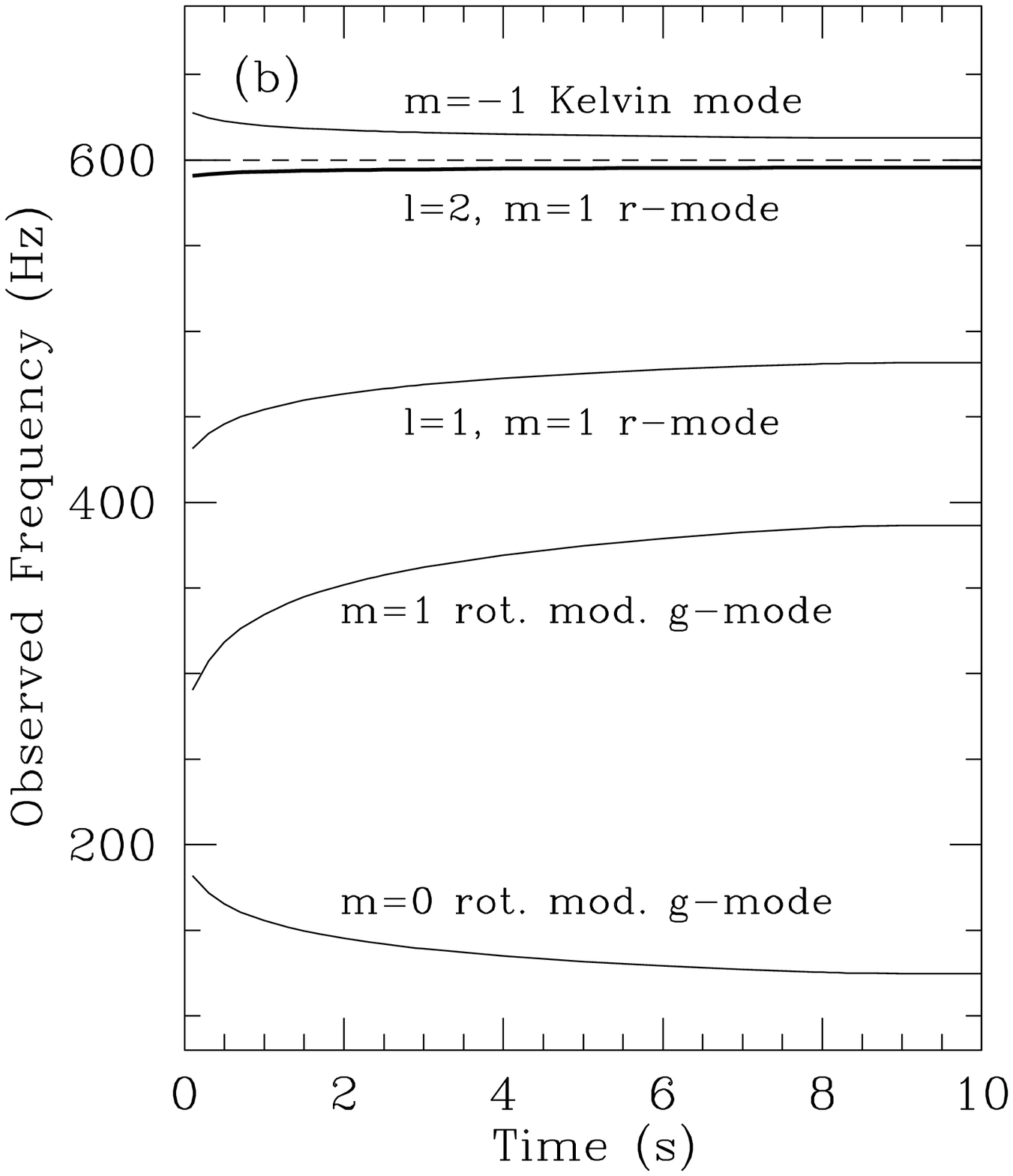}
\caption{Frequency evolution of the lowest angular-order, rotationally
modified modes from Model 1. All modes are $l=1$, unless otherwise noted,
where $l$ refers to the latitudinal quantum number in the slowly spinning
limit. In every case the radial structure is the single-node mode as we
focus on for the burst oscillation. The mode frequencies are denoted by solid
lines, with the NS spin as a dashed line. In ({\it a}) we consider a spin of
$300\ {\rm Hz}$ and find that its frequency shifts are somewhat smaller
than for a $600\ {\rm Hz}$ spin as shown in ({\it b}). Even so, all the frequency
shifts are much larger than the $l=2$, $m=1$ {\it r}-mode that we favor as
the burst oscillations (which is also plotted for comparison; {\it thick solid line}).}
\label{fig:othermodes}
\end{figure*}

  In \S \ref{sec:rotation} we considered the multitude
of rotationally modified solutions available, all with the same single node
radial structure, and identified only one that satisfied the properties of burst
oscillations. The other low angular-order modes resulted in frequency shifts
that were too large, and sometimes in the wrong direction.
Nevertheless, these modes may be compelling to study since
a priori we do not have a reason to exclude them, and more sensitive
observations may reveal these additional modes. Such a measurement
of multiple modes would revolutionize our knowledge of an accreting
NS's mass, radius, spin, and surface layers.

  We calculate the time dependent frequencies from Model 1, this time
considering
a variety of the low angular-order modes, all with the same single-node radial
eigenfunction as used for the burst oscillation. We compare NS spins
of both $300\ {\rm Hz}$ (Fig. \ref{fig:othermodes}a) and $600\ {\rm Hz}$\
(Fig. \ref{fig:othermodes}b) ({\it dashed lines}). Generally, a faster
spin shows larger frequency drifts for these modes because
$\lambda\propto q^2$. This is not the case for the $l=2$, $m=1$ {\it r}-mode
we favor as the burst oscillation, which shows no dependence on $\Omega$
({\it thick solid lines}). When the cooling is fastest at early times, the frequency
shifts are dramatic, ranging anywhere from $\approx15\ {\rm Hz}$ (for the Kelvin
modes) to $\approx100\ {\rm Hz}$ (for the rotationally modified {\it g}-modes).
Such large shifts have not been seen in previous frequency searches of X-ray
burst lightcurves, but may be lurking, undetected. On the other hand, because
of the late time frequency stability afforded by the crustal mode, one could
argue that if these modes existed they would have already been detected
when the frequency evolution is weak. The resolution of this question
depends on the actual oscillation amplitude. If the burst
oscillations are any indication, the amplitudes are largest soon after the
burst peak when the frequency shift is greatest. The stable part of the
frequency might then have an amplitude below detectable levels. It
remains an important exercise to search the extensive
database of burst lightcurves for signatures of oscillations
similar to what we describe here.


\section{Conclusions and Discussion}
\label{sec:concl}

  We considered nonradial surface oscillations as a possible explanation
for the oscillations observed during type I X-ray bursts on non-pulsar
accreting NSs. In studying the time evolution of a shallow surface wave,
we found that this mode changes into a crustal interface wave as the
surface cools, a new and previously unexplored result. This phenomenon allows
us to match both the observed frequency shifts and the high stability of
the asymptotic frequency of burst oscillations. To find frequencies in the
observed range, the modes must be highly modified by rotation,
resulting in the NS spin occurring merely $\approx4\ {\rm Hz}$ above the burst
oscillation frequency. Following similar reasoning to \citet{hey04}, we find
that the majority of the rotationally modified modes
cannot match all of the observed properties of burst oscillations,
because they predict too large of drifts and/or drifts of the wrong direction.
The $l=2$, $m=1$ buoyant {\it r}-mode ($\lambda\approx0.11$) overcomes
all of these difficulties, and for this reason it is the favored mode
to explain the burst oscillations. Perhaps not coincidentally,
this mode has other attractive qualities such as a wide latitudinal eigenfunction
that may make it easier to observe.

  An important mystery that we did not resolve here is why burst
oscillations in the non-pulsars are only seen in the short
$\sim2-10\ {\rm s}$ bursts (not counting the superbursts), a decay timescale
indicative of He-rich fuel at ignition. While Models 1 and 2 show
frequency and amplitude evolution similar to what has been observed for burst
oscillations, Model 3 does not. This is not unexpected because the first two
(especially Model 1) reflect what we expect for He-rich bursts, but it
is at the same time perplexing because we have no reason a priori to explain
why burst oscillations as from Model 3 are not observed.
There must then be some physical mechanism for
why they are not seen. In the study by \citet{cb00}, they find
that modulations in the flux at the burning depth can be
washed out if the thermal time is too long in the hydrostatically
expanding  and contracting surface layers. This effect is dependent
on the mean molecular weight so that it is weak for He-rich bursts, but
fairly strong for mixed H/He bursts. It is possible that a similar phenomena
occurs just above the nonradial modes that we study here.

  There remain many properties of burst oscillations that are not studied
in our work here \citep[for example, see the observations of][]{moc02,moc03}.
The characteristics of the oscillating light curves, the energy dependence of
the pulsed amplitudes, and the phase shifts of different energy components
all can be used to learn about the accreting NSs that are exhibiting these burst
oscillations. Such investigations have just been initiated by \citet{hey05} and
\citet{ls05}.

  The most exciting implication of our study is that the asymptotic
frequencies and drifts are directly related to properties of the surface
NS layers. The crustal interface wave depends on the attributes of the
crust, which in turn is a product of the
X-ray burst properties, and especially its superburst properties.
Superbursts are deep enough to directly impact the composition of the
NS ocean, and eventually,
as material advects down with the accretion flow, the crust. The accreting
low mass X-ray binary 4U $1636-536$ has exhibited oscillations during
a superburst \citep{sm02}, and during X-ray bursts both before and after
(Zhang et al. 1997; M. Muno 2004, private communication),
all with the same asymptotic frequency.
Energetically, superbursts could easily excite crustal waves since they are
roughly a thousand times more energetic than normal X-ray bursts.
For the frequency to be the same in each occurrence the crustal
properties could not have changed during this time. This is possible as long as
4U $1636-53$'s superburst recurrence time is short enough that
the crust keeps a steady state composition.
A crust at $\approx10^{13}\ {\rm g\ cm^{-2}}$ requires
$\approx7\ {\rm yrs}$ at an accretion rate of $10^{-8}\ M_\odot\ {\rm yr^{-1}}$
to be replaced by new, accreted material. Such a timescale favors a crust
that has not changed significantly given that 4U $1636-53$ has produced
three superbursts in the last $4.7\ {\rm yrs}$ \citep{wij01,sm02,kuu04}.

\acknowledgements

  We thank Ed Brown for generously providing us with his NS crust models,
Andrew Cumming for advice on solving the heat diffusion equation, and
Mike Muno for answering our many questions about X-ray burst observations.
Our study has also greatly benefited from the input of a number of people
including Phil Arras, Deepto Chakrabarty, Philip Chang, Anatoly Spitkovsky,
Tod Strohmayer, and Greg Ushomirsky.
This work was supported by the National Science Foundation
under grants PHY99-07949 and AST02-05956, and by the Joint Institute for
Nuclear Astrophysics through NSF grant PHY02-16783.


\vspace{0.75cm}

{\it Note added in proof.}---It has been brought to our attention by Dong Lai that our arguments in
\S \ref{sec:timedependent} for adiabatic evolution through the avoided crossing may not be correct.
We currently use the condition that adiabatic evolution occurs as long as the mode frequency
is much larger than the cooling rate. If the avoided crossing instead acts analogously to
neutrino oscillations (J. Bahcall 1989, Neutrino Astrophysics [Cambridge: Cambridge Univ. Press])
or photon propagation on magnetized neutron stars (D. Lai \& W. C. G. Ho 2002, \apj, 566, 373) the
correct comparison would be the frequency {\it difference} at the avoided crossing versus the cooling
rate. If this is indeed the case, then the surface mode will not solely couple to the crustal interface mode.
Instead, some energy would remain in the surface mode as the frequencies cross. We are currently
studying this problem.

\clearpage
\begin{deluxetable}{l c c c c c c}
  \tablecolumns{5} \tablewidth{0pt}
  \tablecaption{Properties of X-ray Burst Oscillations from Non-pulsars}

  \tablehead{
    \colhead{}
        & \colhead{$\omega/(2\pi)$\tablenotemark{a}}
        & \colhead{$\Delta\omega/(2\pi)$\tablenotemark{b}}
        & \colhead{$\tau$\tablenotemark{c}}
        & \colhead{Radius}
        & \colhead{$L_{\rm pers}/L_{\rm edd}$\tablenotemark{e}}
        & \colhead{References} \\
        \colhead{Object}& \colhead{(Hz)}
        & \colhead{(Hz)}
        & \colhead{(s)}
        & \colhead{Expansion?\tablenotemark{d}}
        & \colhead{(\%)}
        &\colhead{} }
  \startdata
  EXO $0748-676$ & 45 & . . . & . . . & No & $0.9-3.6$&1,2\\
  4U $1916-053$ & 270 & $3.58\pm0.41$ & . . . & No &$1.3-5.6$&3,4 \\
  4U $1702-429$ & 330 & $1.6-3.0$ & $1.9-4.0$ & No & $3-6$&5,6,7,8\\
  & & & & & $1.1-8$ &9\\
  4U $1728-34$ & 363 & $2.1-3.75$ & $1.8-6.2$ & Yes and No & $3.3-9.8$&2,6,7,10,11 \\
  & & & & & $2-7$&8\\
  SAX J$1748.9-2021$ & 410 & . . . & 13.0 & No & . . . &12\\
  KS $1731-260$ & 524 & $0.9-2.0$ & $2.6-4.1$ & Yes &$25$ &7,13,14,8\\
  & & & & & $1.2-14.4$&9\\
  Aql X-1 ($1908+005$) & 549 & $2.4-4.0$ & $2.7$ & Yes and No & $1-3$& 7,15,16,8 \\
  4U $1658-298$ & 567 & $\approx0.5-5$ & $6.9-12.5$ & Yes &$0.2-1$&7,17,18,19 \\
  4U $1636-536$ & 582 & $1.0-3.0$ & . . . & Yes and No & $2.2-7.8$&2,7,20,21,22 \\
  & & & & & $10-15$&8\\
  & & & & & $5.6-12.8$&9\\
  Galactic Center Source & 589 & $\approx1$ & . . . & Yes & . . . & 23\\
  SAX J$1750.8-2980$ &601& $\approx3\tablenotemark{f}$ & $\approx3$ & Yes & $0.2-4.8$& 24,25 \\
  4U $1608-522$ & 619 & $\approx1$ & . . . & Yes & $4.4-17.4$&2,7,26\\
  & & & & & $0.7-6$&8\enddata
\tablenotetext{a}{Asymptotic frequency.}
\tablenotetext{b}{The range of frequency drifts seen from each object.}
\tablenotetext{c}{Estimate of the decay time of cooling in those bursts that exhibit oscillations. This should be considered as only a rough estimate since different authors use different methods of measuring this quantity. The main conclusion to be inferred here is that these are all short, He-like bursts.}
\tablenotetext{d}{Whether or not radius expansions are observed in those bursts that exhibit oscillations. Aql X-1 usually shows radius expansions with burst oscillations, but there is one case when it did not \citep[see][]{mgc04}. }
\tablenotetext{e}{The persistent luminosity as a percentage of the Eddington luminosity. Each range of measurements is shown for those objects with multiple observations. We use $L_{\rm Edd}=2.5\times10^{38}\ {\rm erg\ s^{-1}}$ to be consistent with \citet{for00}.}
\tablenotetext{f}{This drift could be interpreted as preceding the burst peak.}
\tablerefs{(1)\citet{vs04}; (2) \citet{vpl88}; (3) \citet{gal01};
(4) \citet{bor00}; (5) \citet{mss99}; (6) \citet{sm99};
(7) \citet{mun02}; (8) \citet{for00}; (9) \citet{cor03};
(10) \citet{str96}; (11) \citet{vs01}; (12) \citet{kaa03};
(13) \citet{smb97}; (14) \citet{mun00}, (15) \citet{zha98};
(12) \citet{cb00}; (13) \citet{wsf01}; (18) \citet{wij02};
(19) \citet{wij03};
(20) \citet{str98}; (21) \citet{mil00}; (22) \citet{str99};
(23) \citet{str97}; (24) \citet{kaa02};
(25) \citet{nat99};
(26) \citet{har03} }
\end{deluxetable}

\begin{deluxetable}{c c c c c c c}
  \tablecolumns{5} \tablewidth{0pt}
  \tablecaption{Cooling Neutron Star Models}

  \tablehead{
    \colhead{Model} & \colhead{Bursting Layer}
        & \colhead{Ocean/Crust}
        & \colhead{$F_c$ (erg cm$^{-2}$ s$^{-1}$)}
        & \colhead{$\mu_0\equiv\mu/P$}
        & \colhead{$\rho_c$ (g cm$^{-3}$)}
        & \colhead{$\mu_b/\mu_c$} }
  \startdata
  1 & $^{40}$Ca & $^{64}$Zn & $10^{21}$ & $8.8\times10^{-3}$ & $1.8\times10^{9}$ &0.94\\
  2 & $^{40}$Ca & $^{64}$Zn & $3\times10^{21}$ & $8.8\times10^{-3}$ & $9.3\times10^{9}$ &0.94\\
  3 & $^{64}$Zn & $^{104}$Ru & $10^{21}$ & $1.1\times10^{-2}$ & $3.4\times10^{8}$ &0.90
   \enddata
\end{deluxetable}

\end{document}